# One-Minute Derivation of The Conjugate Gradient Algorithm

**Muhammad Ali Raza Anjum**
Army Public College of Management and Sciences
Rawalpindi, PAKISTAN
e-mail: ali.raza.anjum@apcoms.edu.pk

*Abstract*
*One of the great triumphs in the history of numerical methods was the discovery of the Conjugate Gradient (CG) algorithm. It could solve a symmetric positive-definite system of linear equations of dimension N in exactly N steps. As many practical problems at that time belonged to this category, CG algorithm became rapidly popular. It remains popular even today due to its immense computational power. But despite its amazing computational ability, mathematics of this algorithm is not easy to learn. Lengthy derivations, redundant notations, and over-emphasis on formal presentation make it much difficult for a beginner to master this algorithm. This paper aims to serve as a starting point for such readers. It provides a curt, easy-to-follow but minimalist derivation of the algorithm by keeping the sufficient steps only, maintaining a uniform notation, and focusing entirely on the ease of reader.*

**1. Introduction**

Importance of the Conjugate Gradient (CG) algorithm can hardly be overemphasized. It is everywhere: in optimization theory [1, 2], in adaptive filtering [3, 4], in machine learning [5, 6], in image processing [7, 8] and in many others [9-11]. Though it is popular, the algorithm is not easy to learn. The chief difficulty lies with its derivation. Reasons for this difficulty are as numerous as the reasons for its popularity. Too many approaches, too much variety in notation and over emphasis on form and presentation are just to name a few. Books do not try to resolve these issues. Research papers avoid resolving these issues.

There must be an excellent cause for them to do so but they all essentially miss one point: to convey this fascinating algorithm to the aspiring student in a quick, easy and understandable way. Tutorials have made an attempt in this regard. Though they are scarce as well, two of them are truly remarkable. First one tries to systematically address the above mentioned problems but ends up in a long 64 page document in an attempt to explain "everything" [12]. As a result, derivation is almost impossible to follow. Second one gives a quick explanation to the general philosophy of the algorithm but leaves the derivation entirely to the reader [13].

This paper does not aim to do any of these. It neither attempts to explain the philosophy of algorithm nor "everything" of the algorithm. It will present a quick, easy, and understandable proof of the most difficult part of the algorithm: its mathematical derivation. By following this derivation, readers will acquire at least four benefits. Firstly, they will understand the derivation with minimum background knowledge. Only linear algebra will be required. Secondly, they will grasp it in minimum amount of time. Derivation presented is as short as possible. Thirdly, they will learn it with minimum effort. Onus of making things understandable is on author. Finally, once the derivation is finished, they will be able to implement it in a programming language of their choice. The results will be presented in an easy to program manner.

**2. System Model**

Consider following system of linear equations.

$$Ax = b \qquad (1)$$

$A$ is a symmetric $n \times n$ matrix such that $A^T = A$. $x$ and $b$ are $n \times 1$ vectors. We attempt to solve this system iteratively.

$$x_{i+1} = x_i + \alpha d_i \tag{2}$$

$x_i$ is the estimate computed in the $i$-th iteration. $d_i$ is the correction term required in the $i$-th iteration. $\alpha$ is the step-size.

### 3. Computation of Residues

Subtracting the actual solution $x$ from both sides of Eq. (1),

$$x_{i+1} - x = x_i - x + \alpha d_i$$

This results in,

$$e_{i+1} = e_i + \alpha d_i \tag{3}$$

$e_i$ is the error in the $i$-th iteration. At the startup, we make a guess for $x_i$. Since our guess for $x_i$ is purely arbitrary, there will be a residue $r_i$.

$$r_i = b - Ax_i = Ax - Ax_i = A(x - x_i) = -Ae_i$$

Or,

$$r_i = -Ae_i \tag{4}$$

Multiplying both sides of Eq. (3) with $A$,

$$Ae_{i+1} = Ae_i + \alpha Ad_i \tag{5}$$

Replacing the result of Eq. (4) in Eq. (5),

$$-r_{i+1} = -r_i + \alpha Ad_i \tag{6}$$

Or,

$$r_{i+1} = r_i - \alpha Ad_i \tag{7}$$

### 3. First Major Assumption

Now we multiply both sides of Eq. (7) with $r_i^T$.

$$r_i^T r_{i+1} = r_i^T r_i - \alpha r_i^T Ad_i \tag{8}$$

We choose value for $\alpha$ for which,

$$r_i^T r_{i+1} = 0 \tag{9}$$

This is the first crucial decision we make for CG algorithm. So Eq. (8) becomes,

$$0 = r_i^T r_i - \alpha r_i^T Ad_i \tag{10}$$

Finally for $\alpha_i$,

$$\alpha_i = \frac{r_i^T r_i}{r_i^T Ad_i} \tag{11}$$

Hence, Eqs. (2) and (7) become,

$$x_{i+1} = x_i + \alpha_i d_i \tag{12}$$

And,

$$r_{i+1} = r_i - \alpha_i Ad_i \tag{13}$$

### 4. Second Major Assumption

So far we have Eqs. (7) and (11) for the computation of $r_{i+1}$ and the $d_i$ respectively. Now we want to derive the $d_{i+1}$ from $r_{i+1}$ and $d_i$. Let us define the following relationship between all three.

$$d_{i+1} = r_{i+1} + \beta_{i+1} d_i \tag{14}$$

Note that this relationship is quite similar to the pattern in Eq. (12). Now, we multiply both sides of Eq. (14) with $A$.

$$Ad_{i+1} = Ar_{i+1} + \beta_{i+1}Ad_i \tag{15}$$

Further multiplying both sides of Eq. (15) with $d_i^T$,

$$d_i^T Ad_{i+1} = d_i^T Ar_{i+1} + \beta_{i+1}d_i^T Ad_i \tag{16}$$

We want to choose $\beta_{i+1}$ for which,

$$d_i^T Ad_{i+1} = 0 \tag{17}$$

This is the second crucial decision for the CG algorithm. So Eq. (16) becomes,

$$0 = d_i^T Ar_{i+1} + \beta_{i+1}d_i^T Ad_i$$

Finally for $\beta_{i+1}$,

$$\beta_{i+1} = -\frac{d_i^T Ar_{i+1}}{d_i^T Ad_i} \tag{18}$$

## 5. Making the Algorithm Recursive

It will be much more efficient if we can compute $\beta_{i+1}$ from $\alpha_i$ because we already have $\alpha_i$ from Eq. (11). For this purpose, we will exploit $d_i^T Ar_{i+1}$ term in Eq. (18). This can be done by recalling Eq. (7) and multiplying it with $r_{i+1}^T$ on both sides.

$$r_{i+1}^T r_{i+1} = r_{i+1}^T r_i - \alpha_i r_{i+1}^T Ad_i \tag{19}$$

Now we see $d_i^T Ar_{i+1}$ term on the right hand side of Eq. (19). Being a scalar term,

$$d_i^T Ar_{i+1} = r_{i+1}^T Ad_i \tag{20}$$

Substituting our assumption from Eq. (9) in Eq. (19),

$$r_{i+1}^T r_{i+1} = 0 - \alpha_i r_{i+1}^T Ad_i \tag{21}$$

Or,

$$r_{i+1}^T Ad_i = -\frac{1}{\alpha_i} r_{i+1}^T r_{i+1} \tag{22}$$

Using Eq. (20) and remembering that $A$ is symmetric,

$$d_i^T Ar_{i+1} = -\frac{1}{\alpha_i} r_{i+1}^T r_{i+1} \tag{23}$$

Substituting the result of Eq. (23) in (18),

$$\beta_{i+1} = -\frac{d_i^T Ar_{i+1}}{d_i^T Ad_i} = \frac{1}{\alpha_i} \frac{r_{i+1}^T r_{i+1}}{d_i^T Ad_i} \tag{24}$$

Finally, $\beta_{i+1}$ in terms of $\alpha_i$,

$$\beta_{i+1} = \frac{1}{\alpha_i} \frac{r_{i+1}^T r_{i+1}}{d_i^T Ad_i} \tag{25}$$

Now we have Eq. (25) to compute $\beta_{i+1}$ from $\alpha_i$. Only other term required to compute Eq. (25) is $d_i$. But we already have $r_i$ and $r_{i+1}$ from Eq. (13). If we can compute $\beta_{i+1}$ from $r$'s only, it will not only make Eq. (25) more efficient but also recursive. For this purpose, only the dominator term in Eq. (17) has to be expressed in terms of $r$'s for this purpose. So we recall (14) and substitute $i$ in place of $i + 1$ in it.

$$d_i = r_i + \beta_i d_{i-1} \tag{26}$$

Multiplying both sides of (26) first with $A$,

$$Ad_i = Ar_i + \beta_i Ad_{i-1} \tag{27}$$

And then multiplying with $d_i^T$,

$$d_i^T Ad_i = d_i^T Ar_i + \beta_i d_i^T Ad_{i-1} \tag{28}$$

Employing the assumption of Eq. (17) and re-arranging,

$$d_i^T Ad_i = d_i^T Ar_i + 0 \tag{29}$$

Or,

$$r_i^T Ad_i = d_i^T Ad_i \tag{30}$$

Substituting Eq. (30) in Eq. (11),

$$\alpha_i = \frac{r_i^T r_i}{r_i^T Ad_i} = \frac{r_i^T r_i}{d_i^T Ad_i} \tag{31}$$

Or,

$$\alpha_i = \frac{r_i^T r_i}{d_i^T Ad_i} \tag{32}$$

Re-arranging Eq. (32),

$$\alpha_i d_i^T Ad_i = r_i^T r_i \tag{33}$$

Eq. (33) fully expresses the denominator of Eq. (25) in terms of $r$'s. Substituting Eq. (33) in Eq. (25),

$$\beta_{i+1} = \frac{1}{\alpha_i} \frac{r_{i+1}^T r_{i+1}}{d_i^T Ad_i} = \frac{r_{i+1}^T r_{i+1}}{r_i^T r_i} \tag{34}$$

Finally,

$$\beta_{i+1} = \frac{r_{i+1}^T r_{i+1}}{r_i^T r_i} \tag{35}$$

Eq. (35) expresses $\beta_{i+1}$ in terms of $r$'s only and this completes our derivation of CG algorithm.

## 6. The Algorithm Itself

Now we recap the results and list them in the form of an algorithm.

1. Make an initial guess $x_0$.
2. Take $d_0 = r_0 = b - Ax_0$.
3. Compute $\alpha_i$ using Eq. (32).

$$\alpha_i = \frac{r_i^T r_i}{d_i^T Ad_i}$$

4. Update $x$ using Eq. (12).

$$x_{i+1} = x_i + \alpha_i d_i$$

5. Update $r$ using Eq. (7).

$$r_{i+1} = r_i - \alpha_i Ad_i$$

6. Determine $\beta$ using Eq. (35).

$$\beta_{i+1} = \frac{r_{i+1}^T r_{i+1}}{r_i^T r_i}$$

7. Update $d$ using Eq. (14).

$$d_{i+1} = r_{i+1} + \beta_{i+1} d_i$$

## 7. Conclusion

An easy, step-by-step but minimalist derivation of the Conjugate Gradient algorithm is provided in this paper. By following this derivation, the readers, regardless of their background, will not only master this fascinating algorithm in no time but will also be able to program it for their own purposes which will provide them a better understanding of their own problems on which they are working.